\documentclass{jpsj3}
\usepackage{txfonts}

\title{Successive magnetic transitions of Ca$_2$CoSi$_2$O$_7$ in high magnetic fields}

\author{Mitsuru \textsc{Akaki}$^1$\thanks{akaki@ahmf.sci.osaka-u.ac.jp}\thanks{Present address: Center for Advanced High Magnetic Field Science, Osaka University, Toyonaka, 560-0043, Japan}, Hideki \textsc{Kuwahara}$^2$, Akira \textsc{Matsuo}$^1$, Koichi \textsc{Kindo}$^1$, and Masashi \textsc{Tokunaga}$^1$}
\inst{$^1$The Institute for Solid State Physics, The University of Tokyo, Kashiwa 277-8581, Japan \\
$^2$Department of Physics, Sophia University, Tokyo 102-8554, Japan \\
} 

\abst{Magnetic and dielectric properties of \aa kermanite Ca$_2$CoSi$_2$O$_7$ single crystals were investigated in pulsed high magnetic fields.
In magnetic fields along the $c$ axis, this material shows a magnetization plateau in a wide range of field below the saturation. 
Magnetization processes for fields along the $a$ and $b$ axes show multiple anomalies but different traces to each other, indicating the breaking of four-fold symmetry. 
Measurements of the magnetoelectric effects exhibit the changes in electric polarization according to the changes in the spin system. 
The experimentally determined quadratic magnetoelectric tensor is consistent with that expected in the crystal symmetry of orthorhombic $P2_12_12$.

}

\begin{document}
\maketitle

\newpage
Magnetoelectric multiferroic materials have attracted both experimental and theoretical interests because their giant magnetoelectric effects are candidates for future applications\cite{ME1,ME2,ME3}.
The mechanism of spin dependent electric polarization is well explained by several microscopic mechanisms.
The most generally accepted model is the spin-current mechanism \cite{Katsura} in which the ferroelectricity originates in a cycloidal spin structure (e.g., TbMnO$_3$ \cite{KimuraN}).
On the other hand, a long period commensurate spin order (e.g., up-up-down-down-type) induces electric polarization through exchange striction \cite{ES}.
This type of ferroelectricity is realized in highly distorted orthorhombic $R$MnO$_3$ compounds ($R$=Y, Ho, ..., Lu) with $E$-type antiferromagnetism \cite{Etype}, DyFeO$_3$ \cite{DyFeO3}, and Ca$_3$(Co,Mn)$_2$O$_6$ \cite{CCMO}.
A third mechanism is a spin-dependent {\it p-d} hybridization mechanism \cite{Arimapd}.
In this mechanism, the hybridization between transition-metal $d$ orbital and ligand $p$ orbitals is affected by the spin orientation relative to the bond through the spin-orbit interaction.
The variation of the hybridization generates a local electric dipole moment, and the sum of the local electric dipole moment is observed as electric polarization.
As typical examples for this mechanism, magnetic-field induced electric polarization in \aa kermanite Ba$_2$CoGe$_2$O$_7$ \cite{YiAPL,MurakawaPRL} and Sr$_2$CoSi$_2$O$_7$ \cite{AkakiPRB} was recently focused.

\AA kermanite materials attract attention also as an interesting spin system having large easy-plane-type anisotropy against the small exchange interaction.
Magnetic properties in Ba$_2$CoGe$_2$O$_7$ has been studied using the following Hamiltonian \cite{TheoBCGO} as, 
\begin{align}
\mathcal{H}=J \sum_{\langle i,j \rangle}(\hat{S}_i^x \hat{S}_j^x +\hat{S}_i^y \hat{S}_j^y) + J_z \sum_{\langle i,j \rangle}\hat{S}_i^z \hat{S}_j^z+ \Lambda \sum_{i}(\hat{S}_i^z)^2 -g\mbox{\boldmath $h$}\sum_{i}\hat{\mbox{\boldmath $S$}}_i+D_z \sum_{\langle i,j \rangle}(\hat{S}_i^x \hat{S}_j^y +\hat{S}_i^y \hat{S}_j^x), \label{eq:Hami}
\end{align}
where $\langle i,j \rangle$ denotes a pair of nearest-neighbor sites. 
The first two terms in the right side represent the exchange interaction, while the third one denotes the single ion anisotropy.
The fourth and fifth terms are Zeeman energy and the Dzyaloshinskii-Moriya interaction, respectively.
The actual parameters in Ba$_2$CoGe$_2$O$_7$ are estimated 
by the studies of the electron spin resonance and far infrared absorption spectra as $J=2.3$~K, $J_z=1.8$~K and $\Lambda=13.4$~K \cite{PencBCGO}, which shows dominant role of single-ion anisotropy.
In addition, owing to the small exchange interaction in these materials, we can fully polarize  the spins in moderately high magnetic field as demonstrated in our previous study on Sr$_2$CoSi$_2$O$_7$ \cite{AkakiPRB}.
In that study, we found the existence of finite electric polarization in the fully spin polarized state indicating that anitiferromagnetism is not necessary to induce electric polarization through the spin system.
Actually, the magnetic field induced electric polarization was observed even at 300~K in this material.
This behavior can in principle be explained as a magnetoelectric response of second order in the magnetic field, which is called the paramagnetoelectric effect \cite{IT}.

In this study, we focused on another \aa kermanite material Ca$_2$CoSi$_2$O$_7$ \cite{AkakiAPL}.
As seen from the schematic \aa kermanite structure illustrated in the inset of Fig.~1, CoO$_4$ and SiO$_4$ tetrahedra are connected at their corners to form two-dimensional layers, with the layers stacking along the $c$ axis with intervening Ca layers.
As the alkali-earth ions become smaller, the lattice constants at room temperature for $a$-axis ($c$-axis) decreases as 8.410\AA~(5.537\AA), 8.031\AA~(5.165\AA), and 7.842\AA~(5.025\AA) for Ba$_2$CoGe$_2$O$_7, $Sr$_2$CoSi$_2$O$_7$, and Ca$_2$CoSi$_2$O$_7$, respectively.
Ca$_2$CoSi$_2$O$_7$ cannot maintain the $P\overline{4}2_1m$ structure at low temperatures and shows transition to the structure having $3\times3\times1$ times larger unit cell with showing prominent thermal hysteresis \cite{P21212}.
Two different structures were proposed for this superstructure \cite{P-4,P21212}, which remain controversial up to the present.
To clarify the effects of this structural modulation on the magnetic and dielectric properties, we studied magnetization and magnetoelectric effects in single crystals of Ca$_2$CoSi$_2$O$_7$.

Single crystalline samples of Ca$_2$CoSi$_2$O$_7$ and Sr$_2$CoSi$_2$O$_7$ were grown using the floating zone method. 
Through the X-ray diffraction measurements at room temperature, we confirmed that the samples had a tetragonal $P\overline{4}2_1m$ structure without any impurity phase. 
All specimens used were checked by means of the X-ray back-reflection Laue technique, and cut along the crystallographic principal axes into platelike shapes.
The temperature dependence of magnetization was measured using a commercial apparatus (Quantum Design, MPMS).
Pulsed magnetic fields up to 75~T were generated using pulse magnets with pulse duration of $4\sim8$~ms at The Institute for Solid State Physics, The University of Tokyo.
The magnetization in pulsed magnetic fields was measured via the induction method, using coaxial pick-up coils. 
The electric polarization induced by magnetic fields was obtained by integration of the polarization current as a function of time. 
In \aa kermanite materials, electric polarization emerges even without poling electric fields, which is a unique feature of their multiferroicity \cite{AkakiAPL,AkakiPRB,MurakawaPRL}.
Therefore, the polarization current was measured without applying a poling electric field.

Figure 1 shows the temperature dependence of magnetization in (a) Ca$_2$CoSi$_2$O$_7$ and (b) Sr$_2$CoSi$_2$O$_7$ at a field of 0.1~T\@.
In both materials, the magnetization parallel to the $a$ axis ($M_a$) shows a steep increase below N\'eel temperature $T_{\rm N}$ (Ca$_2$CoSi$_2$O$_7$: $T_{\rm N}$=5.7~K, Sr$_2$CoSi$_2$O$_7$: $T_{\rm N}$=7~K) \cite{lt25}.
Below $T_{\rm N}$, the nearest spins along the [110] direction are ordered antiferromagnetically. 
The increase of $M_a$ can be ascribed to spin canting caused by Dzyaloshinskii-Moriya interaction.
On the other hand, the magnetization along the $c$ axis ($M_c$) and $M_a$ show significant difference even in the paramagnetic state owing to the large easy-plane-type anisotropy.
In Sr$_2$CoSi$_2$O$_7$, this large anisotropy also shows up in the difference of the saturation field \cite{AkakiPRB}.
The saturation field was 35~T for the field applied along the $c$ axis, while it was 17~T for the field applied perpendicular to the $c$ axis.
Ca$_2$CoSi$_2$O$_7$ shows smaller disparity between $M_a$ and $M_c$ than Sr$_2$CoSi$_2$O$_7$ as shown in Fig.~1(a), indicating the smaller anisotropy in Ca$_2$CoSi$_2$O$_7$ than that of Sr$_2$CoSi$_2$O$_7$.
As shown in Fig.~2, the slope of magnetization curve along the $c$ axis is also large compared with that of Sr$_2$CoSi$_2$O$_7$ [see also Fig.~2(a) in Ref.~\ref{AkakiPRBlabel}].

Figure 2 displays the magnetization curves of Ca$_2$CoSi$_2$O$_7$ at a temperature of 1.4~K for various field directions.
The magnetization curve for magnetic field along the $a$-axis shows nearly linear increase up to 17~T, and shows gradual increase above this field.
On the other hand, the $M_c$ linearly increases up to 18~T, shows a plateau up to 50~T, and then shows a step-like increase at $\sim$62~T\@.
The difference in the initial slopes of $M_a$ and $M_c$ in Ca$_2$CoSi$_2$O$_7$ is considerably smaller than that in Sr$_2$CoSi$_2$O$_7$ [see also Fig.~2(a) in Ref.~\ref{AkakiPRBlabel}] suggesting the smaller $\Lambda$ in the former material.
The emergence of 1/3 and 2/3 magnetization plateaus in the spin systems described by Eq.~(\ref{eq:Hami}) was claimed in Ref.~\ref{Theo3/2Romhlabel} only when the parameters were chosen to express nearly Ising systems different from the present case. 
Since the magnetization plateau is observed only in Ca$_2$CoSi$_2$O$_7$ which has a small $\Lambda$, the observed magnetization plateau cannot be explained by this theoretical result.
The magnetization in the plateau state is about 85~\% of the saturation magnetization.
Since the Ca$_2$CoSi$_2$O$_7$ has 18~Co$^{2+}$ ions in the unit-cell of the superlattice, the total $S_z$ can be reduced from 27 to $\sim$23 in the plateau region.
Magnetization plateaus have been observed in many frustrated spin system.
In the Ca$_2$CoSi$_2$O$_7$, close look into the differential magnetization curve for $M_a$ shows emergence of multiple phase transitions implying the existence of nearly degenerate ordered states.
Phase transition in this field region shows up clearly for magnetization along the [110] direction.
The $M_{110}$ shows a steep increase at 11~T\@.
In the present study, we cannot specify the spin structure in each phase.
In order to understand this magnetic plateau state, further information on the crystal and magnetic structures are required.

Since the magnetoelectric effects sensitively reflect the changes in symmetry, we studied electric polarization in Ca$_2$CoSi$_2$O$_7$.
Figure 3 shows the magnetic field dependence of (a) magnetization and (b) electric polarization in Ca$_2$CoSi$_2$O$_7$ crystal at various temperatures for magnetic field applied along the [110] direction ($H_{110}$).
At 1.4~K, the electric polarization along the $c$ axis ($P_c$) changed rapidly in low fields, and reached a minimum at 6~T\@.
With increasing fields further, the electric polarization is rapidly reversed its sign accompanied by the magnetization jump at 11~T\@.
This magnetization jump was not observed in other \aa kermanite materials \cite{AkakiPRB,MurakawaPRL}.
At high temperatures, $P_c$ changed in proportional to the square of $H_{110}$.
This behavior can be explained by the paramagnetoelectric effect \cite{AkakiPRB}.
From a microscopic point of view, the electric polarization is caused by a field-induced magnetic moment in the paramagnetic state via the {\it p-d} hybridization mechanism.

Next, we focused on the correlation between paramagnetoelectric effect and structural transition in Ca$_2$CoSi$_2$O$_7$.
According to the group theory, the induced electric polarization $P_i$ in paramagnetic state is defined by the nonzero components of the paramagnetoelectric tensor $\beta_{ijk}$:
\begin{equation}
P_{i}=\sum_{j,k}\frac{1}{2}\beta_{ijk}H_{j}H_{k}.
\end{equation}
Here $i$, $j$ and $k$ represent Cartesian coordinates.
In the case of the point group $\overline{4}2m$, the induced $P_c$ with applying the magnetic field along the [110] direction is given by 
\begin{equation}
P_c=\frac{1}{2}\beta_{312}{H_{110}}^{2}.
\end{equation}
In fact, $P_c$ proportional to the square of $H_{110}$ has been observed as seen from Fig.~4 inset. 
The value of $\beta_{ijk}$ will depend on the crystal structure.
Therefore, we investigated the temperature dependence of $\beta_{312}$ around the structural transition temperature.
Figure 4 shows the temperature dependence of $\beta_{312}$ estimated by $H_{110}$ dependence of $P_c$.
Ca$_2$CoSi$_2$O$_7$ shows the incommensurate-commensurate structural transition accompanied by the thermal hysteresis\cite{P21212}.
According to the previous studies \cite{moduCa,P-4,P21212}, Ca$_2$CoSi$_2$O$_7$ has the ($3\times3\times1$) supercell which is accompanied by the distortion and rotation of CoO$_4$ tetrahedrons below 240~K\@.
As shown in Fig.~4, we have succeeded in observation of the variation in $\beta_{312}$ caused by the structural transition.

Furthermore, we tried the determination of crystal symmetry above the transition temperature by the measurement of paramagnetoelectric tensor.
According to the previous studies \cite{P-4,P21212}, two possible space groups, tetragonal $P\overline{4}$ and orthorhombic $P2_12_12$, have been reported as the low temperature crystal structure of Ca$_2$CoSi$_2$O$_7$.
In the case of $P2_12_12$ (point group is $222$), the induced $P_c$ by paramagnetoelectric effect with the applying the in-plane magnetic field [$H=(H\cos\phi,H\sin\phi,0)$] is expressed as
\begin{align}
P_c&=\frac{1}{2}\beta_{312}H^2\sin 2\phi, \label{eq:P21212}
\end{align}
where $\phi$ is the azimuthal angle of the magnetic field (see Fig.~5 inset).
On the other hand, when the symmetry is $P\overline{4}$ (point group is $\overline{4}$), 
$P_c$ estimated by the paramagnetoelectric tensors is
\begin{align}
P_c=\frac{1}{2}\sqrt{{\beta_{311}}^2+{\beta_{312}}^2}H^2\sin(2\phi+\alpha),\hspace{0.3cm} \alpha =\tan ^{-1}\frac{\beta_{311}}{\beta_{312}}. \label{eq:P-4}
\end{align}
In the case of $P\overline{4}$, since the $\beta_{311}$ is nonzero, finite phase shift $\alpha$ emerges.
Therefore, we tried the determination of crystal symmetry through the measurements of the temperature dependence of $\alpha$.
In order to observe the change of $\alpha$, the $\phi$ dependence of $P_c$ were investigated.
Figure~5 displays the $P_c$ as a function of the magnetic field applied along various directions at 200~K\@. 
The inset shows the $\phi$ dependence of $P_c$ in 35~T at 200~K and 300~K\@.
At 300~K, it has been reported that the crystal symmetry of Ca$_2$CoSi$_2$O$_7$ is $P\overline{4}2_1m$ (point group is $\overline{4}2m$).
In the case of $P\overline{4}2_1m$, the paramagnetoelectric tensor is the same with that of $P2_12_12$.
As shown in the inset, no phase shift is observed between the two temperatures.
The magnitude of $\beta_{311}$ at 200~K estimated from this experiment is about 1~\% of $\beta_{312}$, that is, $\beta_{311}$ can be regarded to be zero in the present accuracy.
This result suggests that the low temperature symmetry in Ca$_2$CoSi$_2$O$_7$ is an orthorhombic $P2_12_12$.

Finally, to check the possible breaking of the four-fold symmetry in the low temperature phase, we performed magnetization measurements along another "$a$-axis" in the tetragonal notation (90 degree different field orientation from the former experiment).
The magnetization labeled $M_b$ shows a different profile from that of $M_a$ as shown in the differential magnetization curve in the inset of Fig.~2.
Therefore, our high-field experiments suggest the orthorhombic $P2_12_12$ rather than tetragonal $P\overline{4}$ for the symmetry in the low temperature phase in Ca$_2$CoSi$_2$O$_7$.
We hope that the observed various magnetic states especially in the plateau region will be clarified by future experimental and theoretical studies based on this $P2_12_12$ structure.

In summary, the magnetic and dielectric properties of Ca$_2$CoSi$_2$O$_7$ single crystals were investigated in pulsed high magnetic fields.
Ca$_2$CoSi$_2$O$_7$ has a small single ion anisotropy.
The magnetization curves show anomalous behaviors.
The magnetization along the [110] direction shows the magnetization jump and the electric polarization parallel to the $c$ axis flips simultaneously.
The magnetization curve for field along the $c$ axis shows the magnetic plateau from 18 to 50~T\@. 
With applying the magnetic fields along the $a$ and $b$ axes, complicated magnetization processes is observed.
Further investigations are necessary to understand the details of several spin states.
We have observed the variation of paramagnetoelectric tensor caused by a structural transition.
By the measurement of paramagnetoelectric tensor, we suggest the low temperature symmetry in Ca$_2$CoSi$_2$O$_7$ is an orthorhombic $P2_12_12$.
Our experimental results show that the electric polarization measurement using a pulsed high magnetic field is very useful in studies of the spin system and crystal structure.

\begin{acknowledgment}
This work was supported by JSPS KAKENHI Grant Number 25800189, 25610087 and 24540383.
\end{acknowledgment}

\newpage
\begin{figure}
\begin{center}
\includegraphics[width=0.5\textwidth, clip]{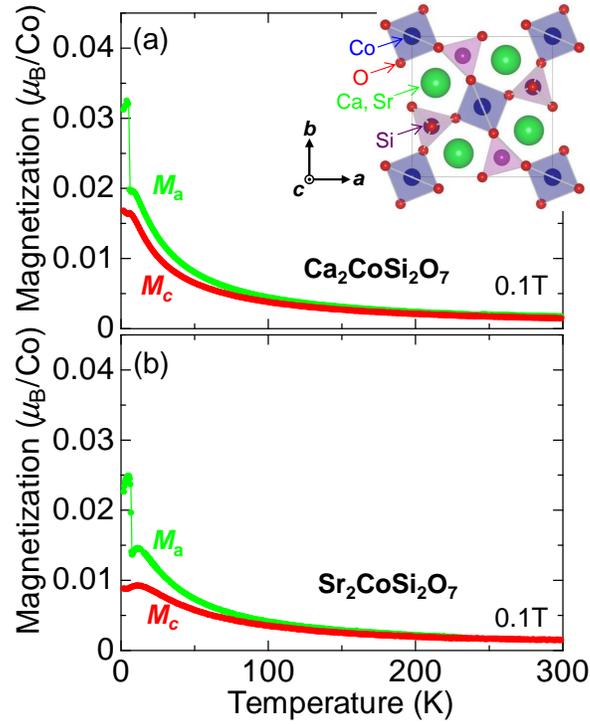}
\vspace{-9pt}
\caption{\label{fig1} (Color online) Temperature dependence of magnetization in (a) Ca$_2$CoSi$_2$O$_7$ and (b) Sr$_2$CoSi$_2$O$_7$ at a field of 0.1~T\@. 
The inset shows schematic crystal structure of \aa kermanite projected onto the $ab$ plane.}
\end{center}
\end{figure}

\begin{figure}
\begin{center}
\includegraphics[width=0.5\textwidth, clip]{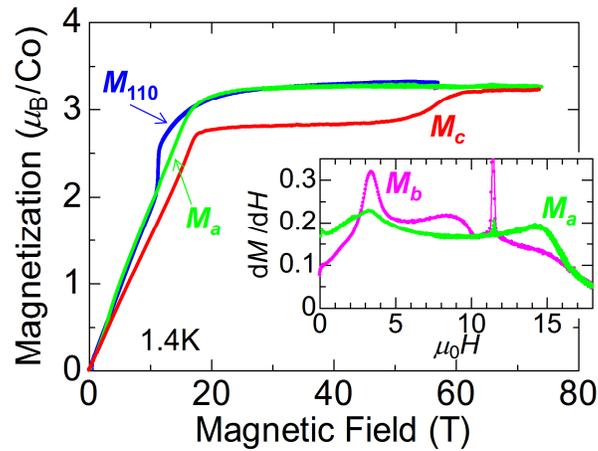}
\vspace{-9pt}
\caption{\label{fig2} (Color online) Magnetization along the each principal axis and [110] direction of the Ca$_2$CoSi$_2$O$_7$ crystal as a function of external magnetic field at a temperature of 1.4~K\@. 
The inset shows the field derivative of the magnetization for the in the field-increasing process.}
\end{center}
\end{figure}

\begin{figure}
\begin{center}
\includegraphics[width=0.5\textwidth, clip]{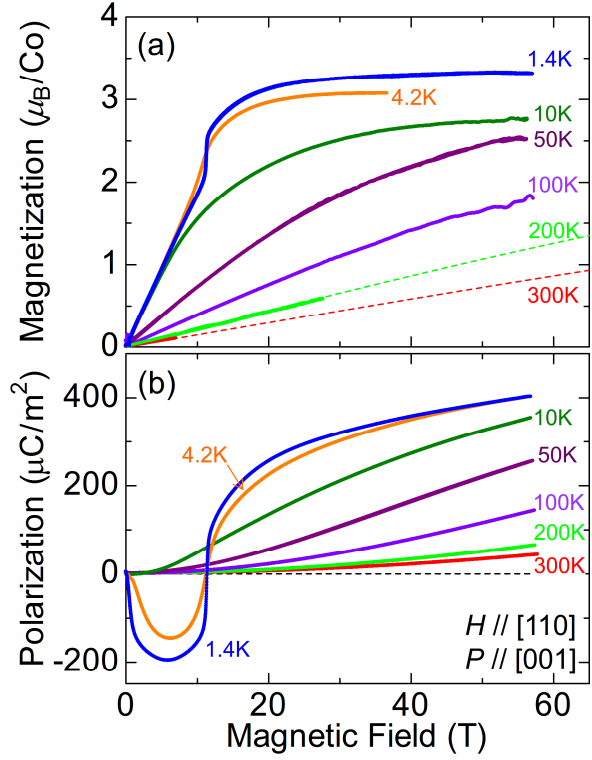}
\vspace{-9pt}
\caption{\label{fig3} (Color online) Magnetic field dependence of (a) magnetization and (b) electric polarization in the Ca$_2$CoSi$_2$O$_7$ crystal at various temperatures.
The broken lines in (a) represent the paramagnetic magnetization curve which is given by the Brillouin function.}
\end{center}
\end{figure}

\begin{figure}
\begin{center}
\includegraphics[width=0.5\textwidth, clip]{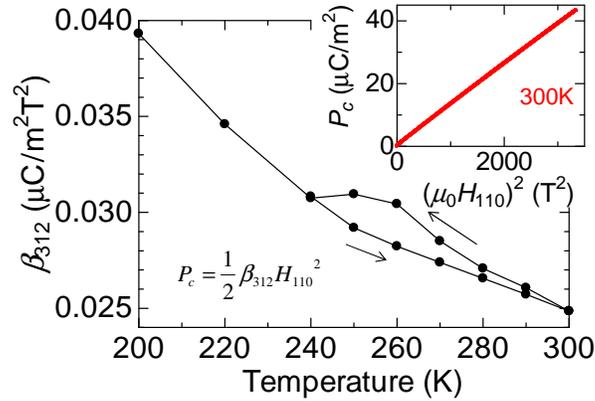}
\vspace{-9pt}
\caption{\label{fig4} (Color online) Temperature dependence of the paramagentoelectric tensor $\beta_{312}$ in the Ca$_2$CoSi$_2$O$_7$ crystal.
The inset shows the electric polarization as a function of the square of the magnetic field at 300~K\@.}
\end{center}
\end{figure}

\begin{figure}
\begin{center}
\includegraphics[width=0.5\textwidth, clip]{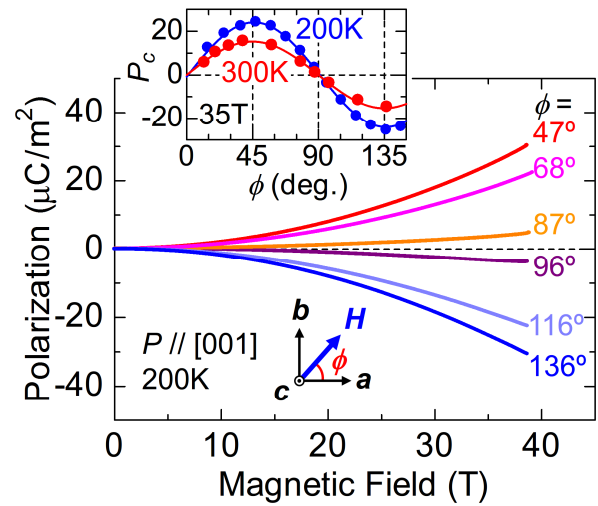}
\vspace{-9pt}
\caption{\label{fig5} (Color online) Electric polarization along the $c$ axis of the Ca$_2$CoSi$_2$O$_7$ crystal as a function of magnetic field applied along various directions at 200~K\@. 
The inset shows the magnetic field angle $\phi$ dependence of the electric polarization at 35~T\@.
The angle $\phi$ is defined as the angle between [100] and the $H$ direction in the (001) plane.}
\end{center}
\end{figure}

\end{document}